\documentclass[aip,graphicx,amsmath,amssymb,preprint]{revtex4-1}

\usepackage{float}
\usepackage{graphicx}
\usepackage[justification=justified]{caption}

\usepackage{subcaption}
\usepackage{hyperref}
\usepackage{amsmath}
\usepackage[capitalize]{cleveref}
\usepackage{xcolor}
\usepackage{array}
\usepackage{amssymb}
\usepackage{adjustbox}

\newcommand\R[1]{\textcolor{black}{#1}}

\begin{document}
\title{An Uncertainty Visualization Framework for Large-Scale Cardiovascular Flow Simulations: A Case Study on Aortic Stenosis}
%
%
\author{Xiao Xue}
\affiliation{ 
Centre for Computational Science, University College London, London, UK
}
\author{Tushar M. Athawale}
\affiliation{Oak Ridge National Laboratory, Oak Ridge, TN, USA}
\author{Jon W. S. McCullough}
\affiliation{ 
Centre for Computational Science, University College London, London, UK
}
\affiliation{ 
School of Mechanical and Aerospace Engineering, Queen?s University Belfast, Belfast, UK
}
\author{Sharp C. Y. Lo}
\affiliation{ 
Centre for Computational Science, University College London, London, UK
}
\author{Ioannis Zacharoudiou}
\affiliation{ 
Centre for Computational Science, University College London, London, UK
}
\author{B\'alint Jo\'o}
\affiliation{Oak Ridge National Laboratory, Oak Ridge, TN, USA}
\affiliation{NVIDIA Corporation, Santa Clara, USA}
\author{Antigoni Georgiadou}
\affiliation{Oak Ridge National Laboratory, Oak Ridge, TN, USA}
\author{Peter V. Coveney}
\email{p.v.coveney@ucl.ac.uk}
\affiliation{Centre for Computational Science, University College London, London, UK}
\affiliation{Advanced Research Computing Centre, University College London, London, UK}
\affiliation{Informatics Institute, Faculty of Science, University of Amsterdam, Amsterdam, Netherlands}
              
\begin{abstract}
We present a generalizable uncertainty quantification (UQ) and visualization framework for lattice Boltzmann method simulations of high Reynolds number vascular flows, demonstrated on a patient-specific stenosed aorta. The framework combines EasyVVUQ for parameter sampling with large-eddy simulation turbulence modeling in HemeLB, and executes ensembles on the Frontier exascale supercomputer. Spatially resolved metrics, including entropy and isosurface-crossing probability, are used to map uncertainty in pressure and wall shear stress fields directly onto vascular geometries. Two sources of model variability are examined: inlet peak velocity and the Smagorinsky constant. Inlet velocity variation produces high uncertainty downstream of the stenosis where turbulence develops, while upstream regions remain stable. Smagorinsky constant variation has little effect on the large-scale pressure field but increases WSS uncertainty in localized high-shear regions. In both cases, the stenotic throat manifests low entropy, indicative of robust identification of elevated WSS. By linking quantitative UQ measures to three-dimensional anatomy, the framework improves interpretability over conventional 1D UQ plots and supports clinically relevant decision-making, with broad applicability to vascular flow problems requiring both accuracy and spatial insight.

\keywords{Uncertainty visualization, uncertainty quantification, Large-eddy simulations, lattice Boltzmann method, hemodynamics}
\end{abstract}

\maketitle
\section{Introduction}

Computational fluid dynamics (CFD) plays a pivotal role across a wide range of scientific and engineering disciplines, from understanding turbulent flows in aerodynamics~\cite{biferale2004multifractal,xue2024physics}, through optimizing industrial processes~\cite{norton2006computational}, to modeling physiological systems in biomedical research~\cite{mazzeo2008hemelb}. In cardiovascular applications, in particular, CFD enables non-invasive investigations of complex hemodynamics and supports clinical decision-making through patient-specific simulations~\cite{kim2013patient,groen2018validation}. Among CFD approaches, direct numerical simulation (DNS) provides the highest fidelity by resolving all relevant spatial and temporal scales of the Navier?Stokes equations using numerical schemes such as finite-difference~\cite{strikwerda2004finite}, finite-volume~\cite{eymard2000finite}, and finite-element methods~\cite{reddy1993introduction}. However, the extremely high computational cost of DNS renders it impractical for most real-world biomedical applications?particularly those involving complex vascular geometries and high Reynolds number flows, such as occur in aortic hemodynamics~\cite{xue2024lattice}.

To balance accuracy and computational cost, large-eddy simulation (LES) has emerged as a practical alternative. LES resolves large-scale turbulent structures while modeling the subgrid-scale (SGS) motions~\cite{smagorinsky1963general,spalart1992one}, thereby significantly reducing the cost relative to DNS. Nonetheless, LES still demands fine resolution near walls to capture boundary-layer dynamics accurately~\cite{chapman1979computational,choi2012grid,yang2021grid}, which presents challenges in geometrically complex domains like blood vessels.

An increasingly popular approach to overcoming these limitations, especially in biomedical applications, is the lattice Boltzmann method (LBM)\cite{succi2001lattice}. LBM provides a mesoscopic alternative to traditional CFD by solving the Boltzmann transport equation on a discrete lattice. Its strengths include high computational efficiency, intrinsic parallelizability, and simplified implementation of complex boundary conditions. These features make LBM particularly effective in handling complex geometries and enable its use across scales?from microscale and mesoscale flow applications\cite{xue2018effects,chiappini2018ligament,zacharoudiou2023development,nash2014choice}, to macroscopic flow simulations~\cite{hou1994lattice,karlin1999perfect,lallemand2000theory,shao2022near}. This scalability also makes it well-suited for ensemble simulations, as required in uncertainty quantification (UQ) studies.
HemeLB~\cite{mazzeo2008hemelb} is an open-source LBM-based CFD framework specifically designed for simulating blood flow in sparse and intricate vascular geometries. Optimized for high performance on modern computing platforms, HemeLB has demonstrated success in large-scale patient-specific domains~\cite{afrouzi2020simulation,lo2024uncertainty}, and has become a widely used tool in cerebral~\cite{mccullough2023high} and aortic hemodynamics~\cite{lo2025multi}. Most vascular studies using HemeLB focus on laminar or transitional flow regimes with Reynolds numbers below 1000. However, in pathological conditions such as aortic coarctation, local flow disturbances can lead to regions of high Reynolds number where turbulence modeling becomes necessary. To address this, HemeLB incorporates LES with subgrid-scale models such as Smagorinsky~\cite{smagorinsky1963general,hou1994lattice,xue2022synthetic}. \R{Importantly, HemeLB has demonstrated excellent scaling properties up to and beyond the exascale level, which enabled us to perform the large ensemble simulations reported here on the Frontier supercomputer~\cite{zacharoudiou2023development}.}

The Smagorinsky constant ($C_{\text{smag}}$) plays a pivotal role in controlling the modeled subgrid-scale viscosity, and its choice directly impacts key flow quantities?particularly in regions of high shear stress relevant to clinical diagnostics. This sensitivity underscores the critical importance of UQ in high-Reynolds-number LBM simulations. Variability in model parameters?such as the Smagorinsky constant, inflow velocity profiles, or boundary conditions?can substantially influence hemodynamic predictions like pressure gradients and wall shear stress distributions.

UQ has been widely applied in molecular dynamics~\cite{bhati2018uncertainty,coveney2025molecular,wan2021uncertainty,coveney2021reliability}, turbulence modeling~\cite{tracey2013application,o2024quantifying} and hemodynamics~\cite{chen2013simulation}, providing systematic methods to assess how uncertainties propagate through CFD models. For example, Lo et al.~\cite{lo2024uncertainty} investigated the impact of peripheral arterial disease on abdominal aortic aneurysms using the EasyVVUQ framework~\cite{richardson2020easyvvuq}. However, in the biomedical LBM community, especially for high-Reynolds-number vascular flows, there remains a lack of frameworks that provide spatially interpretable UQ results. Quantifying and visual uncertainty in three-dimensional simulations is particularly challenging due to the nonlinear nature of data transformations, the high computational cost and limited scalability of empirical approaches such as Monte Carlo, and the absence of established techniques for representing uncertainty directly in complex 3D domains. As a result, many studies reduce UQ outputs to one-dimensional summaries, such as error bars or boxplots~\cite{boxplotsTukey}, which can obscure spatial patterns of variability that are important for clinical interpretation. Recently, several advanced visualization approaches have been developed that enable the direct representation of uncertainty in three-dimensional spatial domains~\cite{TA:Johnson:2003:nextStepVisErrors,TA:Kamal:2021:UQvisSurvey,ensembleUQ,TA:Bonneau:2014:StateOftheArtUQ,TA:Potter:2012:UQtaxonomy,TA:Brodlie:2012:RUDV}, offering the potential to link UQ analysis more intuitively to anatomical structures.

In this work, we take an initial step toward filling this gap by introducing a UQ framework tailored for LBM-based hemodynamic simulations. Our focus is on assessing the influence of the Smagorinsky constant and peak inflow velocity on spatially distributed uncertainties in aortic flow predictions. \R{To our knowledge, this study is among the first to demonstrate uncertainty-aware cardiovascular flow simulations successfully completed on a realized exascale system, highlighting the potential of exascale computing for clinical-scale ensemble studies.} Unlike traditional UQ approaches that rely solely on aggregate statistical measures, our framework emphasizes spatial visualization of uncertainty fields mapped onto sparse vascular geometries, enabling direct anatomical interpretation. This spatially resolved approach is particularly valuable in clinical contexts, where practitioners often require explainable, localized insights rather than abstract statistical summaries. By reducing the interpretability barrier, we aim to facilitate broader adoption of UQ-informed simulations in clinical decision-making and future clinical workflows.

The remainder of this work is organized as follows. In Section~\ref{sec:methods}, we provide an overview of the lattice Boltzmann method and detail the implementation of the Smagorinsky large-eddy simulation model within this framework. We also introduce the uncertainty visualization methodology used to explore the uncertainty region of simulation outcomes to key model parameters through spatial visualization. Section~\ref{sec:results} presents validation results, including comparisons between our LES simulations, experimental measurements~\cite{ding2021transitional}, and a canonical DNS benchmark~\cite{kim1987turbulence}. We then apply the UQ framework to examine the impact of variability in the Smagorinsky constant and inflow velocity profiles on the spatial distribution of uncertainties. Finally, Section~\ref{sec:conclusion} summarizes our key findings and discusses implications for future CFD studies of aortic coarctation.

\section{Methods}\label{sec:methods}
\subsection{The lattice Boltzmann method}\label{sec:method-lbm}

Within this study, we make use of HemeLB~\cite{zacharoudiou2023development}, which is specifically optimized for sparse vascular geometries and parallel execution. Its communication patterns and domain decomposition strategy are designed to minimize overhead in highly irregular vascular networks, enabling excellent strong and weak scaling across modern supercomputers, including exascale platforms such as Frontier. We employs a three-dimensional (3D) lattice Boltzmann model using 19 discrete velocity directions, known as the D3Q19 model. Each lattice cell is defined by its spatial position $\mathbf{x}$ and time $t$, and is associated with a discrete velocity set $\mathbf{c}_i$ where $i \in \{0, 1, \ldots, Q-1\}$ and $Q = 19$. The evolution of the distribution functions follows the lattice Boltzmann equation:
\begin{equation}
\label{eq:lbe}
f_i(\mathbf{x}+\mathbf{c}_i \Delta t, t+\Delta t) = f_i(\mathbf{x}, t) - \Omega \left[f_i(\mathbf{x}, t) - f_i^{eq}(\mathbf{x}, t)\right],
\end{equation}
where $\Omega = \frac{\Delta t}{\tau}$ is the Bhatnagar-Gross-Krook (BGK) single-relaxation-time collision operator~\cite{succi2001lattice}, and $\tau$ is the relaxation time. The equilibrium distribution function $f_i^{eq}$ is defined as:
\begin{equation}
\label{eq:local_eq}
f_i^{eq}(\mathbf{x}, t) = w_i \rho(\mathbf{x}, t) \left[1 + \frac{\mathbf{c}_i \cdot \mathbf{u}(\mathbf{x}, t)}{c_s^2} + \frac{(\mathbf{c}_i \cdot \mathbf{u}(\mathbf{x}, t))^2}{2c_s^4} - \frac{\mathbf{u}(\mathbf{x}, t) \cdot \mathbf{u}(\mathbf{x}, t)}{2c_s^2} \right],
\end{equation}
where $w_i$ are the lattice weights ($w_0 = 1/3$, $w_{1-6} = 1/18$, $w_{7-18} = 1/36$), $\rho(\mathbf{x}, t)$ is the fluid density, and $\mathbf{u}(\mathbf{x}, t)$ is the macroscopic velocity. The time step $\Delta t$ is set to unity in lattice units. The kinematic viscosity $\nu$ is given by:
\begin{equation}
\label{eq:nu}
\nu = c_s^2 \left(\tau - \frac{1}{2}\right) \Delta t,
\end{equation}
where the lattice speed of sound satisfies $c_s^2 = 1/3$ in lattice units.

Macroscopic quantities are obtained from moments of the distribution functions:
\begin{equation}
\label{eq:density}
\rho(\mathbf{x}, t) = \sum_{i=0}^{Q-1} f_i(\mathbf{x}, t),
\end{equation}
\begin{equation}
\label{eq:momentum}
\rho(\mathbf{x}, t)\mathbf{u}(\mathbf{x}, t) = \sum_{i=0}^{Q-1} f_i(\mathbf{x}, t)\mathbf{c}_i.
\end{equation}
\subsection{Smagorinsky subgrid-scale modelling}\label{sec:method-sgs}

We employ the Smagorinsky subgrid-scale model within the LBM framework for large-eddy simulation~\cite{smagorinsky1963general,hou1994lattice,koda2015lattice}. The effective viscosity $\nu_{\mathrm{eff}}$ is modeled as the sum of the molecular viscosity $\nu_0$ and a turbulent viscosity $\nu_t$:
\begin{equation}
\label{eq:sgs}
\nu_{\mathrm{eff}} = \nu_0 + \nu_t,
\end{equation}
\begin{equation}
\label{eq:nu_t}
\nu_t = C_{\mathrm{smag}} \Delta^2 \left|\bar{\mathbf{S}}\right|,
\end{equation}
where $\left|\bar{\mathbf{S}}\right|$ is the magnitude of the filtered strain rate tensor, $C_{\mathrm{smag}}$ is the Smagorinsky constant, and $\Delta$ is the filter width, typically taken as the lattice spacing. The effective viscosity from~\cref{eq:sgs} is substituted into~\cref{eq:nu} to incorporate turbulence effects. A more detailed derivation can be found in~\cite{xue2022synthetic}.

\subsection{Sponge layer implementation}\label{sec:sponge-sgs}

To damp velocity fluctuations near the outlet, we implement a sponge layer: a region of gradually increasing viscosity that absorbs disturbances~\cite{guo1994comparison,adams1998direct,lo2025multi}. Within this zone, the kinematic viscosity $\nu_s$ is defined as:
\begin{equation}
\label{eq:sponge_visco}
\nu_s = \nu_{\mathrm{eff}} \left[1 + (p_s - 1) \left(\frac{\|d\|}{w_{\mathrm{width}}}\right)^2\right],
\end{equation}
where $p_s$ is a viscosity amplification factor (set to $p_s=1000$ in this study), $\|d\|$ is the distance from the outlet plane, and $w_{\mathrm{width}}$ is the width of the sponge zone. This technique helps ensure numerical stability and minimizes artificial reflections from domain boundaries.

\subsection{Uncertainty quantification and visualization}
\subsubsection{Parameter sampling}

We use the EasyVVUQ framework~\cite{richardson2020easyvvuq,edeling2021impact} to generate ensembles of simulations for uncertainty quantification. EasyVVUQ provides a non-intrusive interface for defining uncertain model parameters, sampling them according to prescribed probability distributions, and managing large-scale simulation campaigns on HPC platforms.

In this work, uncertain parameters include the inlet peak velocity $v_{\max}$ and the Smagorinsky constant $C_{\mathrm{smag}}$ used in the LES turbulence model. Each parameter is assigned a uniform distribution over its range of interest. Sampling is performed using the polynomial chaos expansion sampler in EasyVVUQ, with the total polynomial order $p_o$ specified by the user based on the desired balance between accuracy and computational cost. The polynomial order determines the number of quadrature nodes $N_q$ through the relation
\[
    N_q = (p_o + 1)^d,
\]
where $d$ is the number of uncertain parameters. \R{While EasyVVUQ generates the sampling plans, the execution and job submission to leadership-class computing facilities are handled through complementary workflow management tools such as FabSim3~\cite{groen2023fabsim3}.}These nodes $\{\xi_j\}$ and associated weights $\{w_j\}$ are selected according to the input distribution; for uniform priors, a Gauss?Legendre quadrature rule is used. The polynomial chaos expansion coefficients $\alpha_k$ are then approximated as
\begin{equation}
    \alpha_k \approx \sum_{j=1}^{N_q} f(\xi_j) \, \Psi_k(\xi_j) \, w_j ,
\end{equation}
where $f(\xi_j)$ is the model output for the $j$-th quadrature node, $\Psi_k$ is the orthonormal polynomial basis function of order $k$, and $\alpha_k$ are the expansion coefficients. The choice of $p$ directly controls the highest polynomial order in $\Psi_k(\xi)$, and hence the accuracy of the surrogate representation.

\subsubsection{Uncertainty visualization}\label{subsec:uqv}
For uncertainty visualization, we employ state-of-the-art three-dimensional uncertainty visualization techniques to assess variability across an ensemble of LBM simulations. The analysis focuses on the positional uncertainty of isosurfaces extracted from clinically relevant fields such as pressure and wall shear stress (WSS). This spatially resolved approach complements conventional statistical summaries by identifying where model predictions are most stable or most sensitive to parameter variation.

The method is based on the probabilistic isosurface framework of Athawale et al.~\cite{TA:Athawale:2021:topoMappingUncertaintyMarchingCubes} and P\"{o}thkow et al.~\cite{pothkow2011probabilistic}, which performs statistical aggregation across ensemble runs to locate uncertain and stable regions in the data. Let the simulation domain be $\mathcal{D} \subset \mathbb{R}^3$ and the ensemble consist of $m$ runs. For each spatial location $\mathbf{x} \in \mathcal{D}$, let $F(\mathbf{x}) \in \mathbb{R}^m$ denote the scalar field values from all ensemble members. The distribution of values at each vertex is modeled as an independent Gaussian $\mathcal{N}(\mu, \sigma^2)$, where $\mu$ and $\sigma$ are the sample mean and standard deviation computed over the ensemble.
Given an isovalue $c$, the probability that the value at vertex $v$ is below $c$ is
\begin{equation}
    \mathrm{Pr}(v^-) = \frac{1}{2} \left[ 1 + \mathrm{erf} \left( \frac{c - \mu}{\sqrt{2}\,\sigma} \right) \right],
\end{equation}
which corresponds to the cumulative distribution function (CDF) of the Gaussian distribution. The probability that the value is above $c$ is then $\mathrm{Pr}(v^+) = 1 - \mathrm{Pr}(v^-)$.

For a given grid cell $q$ with eight vertices $v_0, \dots, v_7$, the signs of the vertices relative to $c$ define one of $2^8 = 256$ possible marching cubes topology cases. The probability of a specific topology is computed as the product of the vertex probabilities. For example, the probability of the case where $v_0$ is negative and all other vertices are positive is
\[
    \mathrm{Pr}(v_0^-) \prod_{i=1}^7 \mathrm{Pr}(v_i^+).
\]
Repeating this for all $256$ cases yields the full topology probability distribution for each cell.

From this distribution, two complementary spatial uncertainty metrics are derived. The Shannon entropy~\cite{TA:Athawale:2021:topoMappingUncertaintyMarchingCubes} is defined as
\begin{equation}
    E(q) = - \sum_{t=1}^{256} p_t(q) \log_2 p_t(q),
\end{equation}
where $p_t(q)$ is the probability of topology case $t$ in cell $q$. Entropy measures the information-theoretic spread of possible topologies; higher values indicate greater variability in isosurface position across the ensemble, while low values indicate stability. The isosurface-crossing probability~\cite{pothkow2011probabilistic} is
\begin{equation}
    I(q) = \sum_{t=2}^{255} p_t(q),
\end{equation}
excluding cases $t=1$ and $t=256$ where the isosurface does not intersect the cell. This metric measures the likelihood that the isosurface intersects the cell in at least one ensemble member.

Both $E(q)$ and $I(q)$ are mapped to volume renderings for visual inspection, with color encodings highlighting regions of high and low uncertainty. By linking these metrics directly to anatomical geometries, the method enables precise, spatially localized interpretation of uncertainty patterns in clinically relevant hemodynamic fields.

\section{Validation of large eddy simulation using HemeLB}\label{sec:results}
In this section, we initially present the configuration of stenosis in turbulent channel flow for validation purposes, and subsequently conduct a comparative analysis with both experimental Particle Image Velocimetry (PIV)  (Ding et al.)~\cite{ding2021transitional} and DNS datasets~\cite{kim1987turbulence}. Subsequently, the potential of the LBM-based LES (LBM-LES) approach is elaborated upon, demonstrating its capability to simulate aortic flow at a resolution of $100 \mu m$, which is coarser than the resolution attainable with the exclusive use of the BGK collision operator.

\begin{figure}[H]
\includegraphics[width=\textwidth]{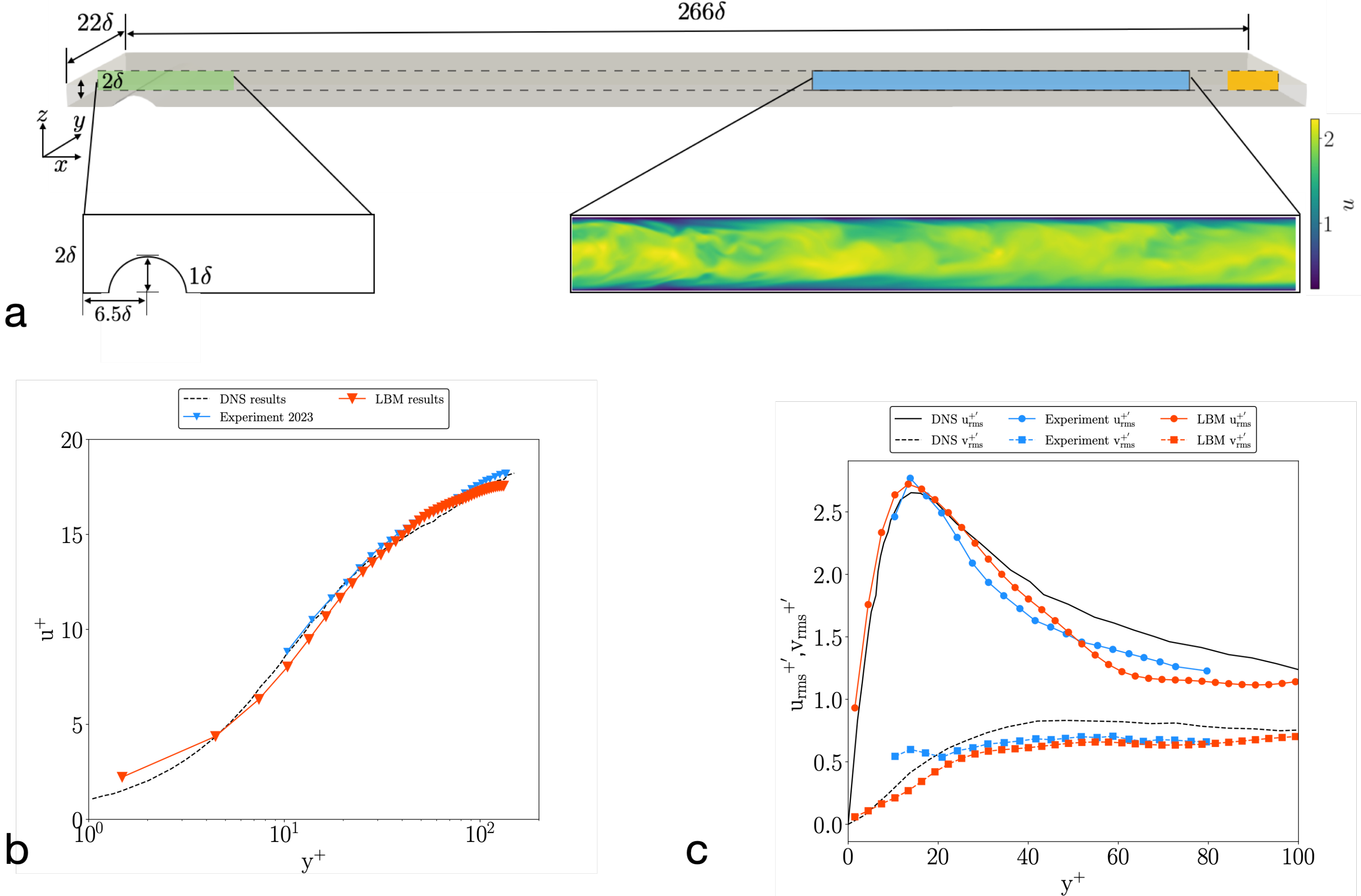}
\caption{Panel a: Illustration of a stenotic channel flow simulation setup: The depicted stenotic channel commences with a semi-cylindrical bump, possessing a height of $1\delta$, obstructing $50\%$ of the channel's cross-section. The green zone demonstrates the semi-cylindrical obstacle's configuration. The blue zone shows a snapshot where the channel flow has evolved into a fully developed turbulent velocity profile. The orange zone represents the sponge zone, designed to absorb reflective waves emanating from the outlet. Panel b: representation of $\mathrm{u^{+}}$ as a function of $y^+$, where the black dotted line illustrates the DNS reference data \cite{kim1987turbulence}. The blue triangles depict the PIV experiment data from Ding et al.\cite{ding2021transitional}. Panel c:
Presentation of $\mathrm{u_{rms}}^{+\prime}$ and $\mathrm{v_{rms}}^{+\prime}$ as functions of $y^+$, with the black and black dotted lines representing the DNS reference data \cite{kim1987turbulence}. The blue circles and squares correspond to $\mathrm{u_{rms}}^{+\prime}$ and $\mathrm{v_{rms}}^{+\prime}$ from the experimental results \cite{ding2021transitional}, respectively. Additionally, the red circles and squares depict the LBM simulation outcomes for $\mathrm{u_{rms}}^{+\prime}$ and $\mathrm{v_{rms}}^{+\prime}$, respectively.} 
\label{figure: sketch}
\end{figure}
\subsection{Validation of the turbulent channel flow simulation}
As a validation exercise for the method, we first consider a high-resolution stenotic channel flow simulation. As depicted in~\cref{figure: sketch}, the dimensions of the stenotic channel flow are configured as $L_{x} \times L_{y} \times L_{z}$, where $L_x = 266\delta$, $L_y = 22\delta$, and $L_z = 2\delta$ correspond to the streamwise, spanwise, and vertical directions, respectively. Here, $\delta$ denotes the turbulent boundary layer thickness, set at $\delta = 0.0045 m$. The lattice resolution is set to $100 \mu m$, resulting in a simulation domain composed of approximately $1.0\times 10^9$ lattice cells. \R{The Smagorinsky constant is set to $C_{\mathrm{smag}} = 0.01$ in this simulation.} The domain is driven by an inlet boundary condition characterized by a plug velocity profile, where the maximum velocity reaches $2.16m/s$. The outlet is configured with a pressure-free boundary condition to facilitate flow egress without additional resistance. To enhance stability, a sponge zone, highlighted in the figure in orange, is positioned prior to the outlet to absorb reflective waves triggered by non-equilibrium bounce-back effects at the outlet. The sponge zone is set to $4\delta$ to ensure the stability of the simulation.
Both the upper and lower planes in the $z$ direction are assigned no-slip boundary conditions.

Unlike the setups described in other studies~\cite{koda2015lattice,xue2022synthetic}, periodic boundary conditions are not employed in the spanwise direction. Instead, a wide channel with no-slip boundary conditions, mirroring the experimental configurations~\cite{ding2021transitional}, is implemented. To accelerate the turbulent transition adaptation length, a semi-cylindrical obstacle with the height of a half-sphere is positioned at $6.5 \delta$ from the inlet. Initially, the \R{inlet flow} is established as laminar, but the presence of the obstacle disrupts the flow's symmetry, leading to a gradual transition to turbulent flow. The \R{inlet velocity} is incrementally increased from $0.1 m/s$ to $2.16 m/s$ over a duration of 1 second. 


Upon reaching the maximum velocity, the simulation continues for an additional second to ensure that the turbulent flow fully develops. The channel flow simulation yields a Reynolds number, \(Re\), approximately equal to 4600. Simultaneously, the friction Reynolds number, \(Re_{\tau}\), is estimated to be around 130. The expressions defining \(Re\) and \(Re_{\tau}\) are presented as
\begin{equation}
\label{eq:re}
    Re = u_{\mathrm{max}}*H/{\nu},\hspace{.6in} Re_{\tau} = u_{\tau}*H/{\nu}
\end{equation}
where $Re$ is characterized by the maximum velocity and $Re_{\tau}$ is measured by the shear velocity. $H$ is the channel flow height which is $9\times10^{-3} m$ and $\nu$ is the kinematic viscosity which is set to $4\times10^{-6} m^2/s$. 
We commence collecting statistics after the 1 million simulation timestep, focusing on the blue region illustrated in~\cref{figure: sketch}, positioned at the channel's mid cross-section, extending from $0.2m$ to $0.4m$. The spatial ensemble average of the streamwise velocity, $u^{+}$, as a function of $y^{+}$, is calculated. These variables represent the dimensionless mean velocity and the dimensionless distance from the wall, respectively, offering insights into the flow characteristics within the specified region.

\begin{equation}
    \label{eq:y_plus}
u^{+}=\frac{\left\langle u \right\rangle }{u_{\tau}},\hspace{.6in} y^{+}=y\frac{\sqrt{\tau_{\omega}/\rho}}{\nu},
\end{equation}
where $\left\langle \cdot \right\rangle$ denotes the ensemble average over the streamwise direction. 
$u$ represents the streamwise velocity, and $u_{\tau}$ is defined as the shear velocity. The variable $y$ denotes the physical distance to the wall, while $\tau_{w}$ represents the wall shear stress. $\nu$ is the kinematic viscosity from the LBM simulation obtained via~\cref{eq:nu}. As depicted in~\cref{figure: sketch} (b), the experimental data begins from $y^+ = 10$, whereas the $y^+$ for the first cell near the wall in the LBM-LES simulation is approximately $y^+ = 1.5$. Although there is a minor deviation in the first cell, the LBM-LES simulation aligns well with both experiment and DNS references. For $y^+>30$, the LBM-LES results deviate slightly from the DNS data but remain in close agreement with the experimental results. Overall, the LBM-LES implementation demonstrates good concordance with both experimental and DNS simulations.

Furthermore, we examined the dimensionless root mean square (RMS) for two velocity components in the streamwise and vertical directions.  $\mathrm{u}^{+\prime}_{\mathrm{rms}}$, $\mathrm{v}^{+\prime}_{\mathrm{rms}}$ are the dimensionless RMS velocity components that are normalized with the shear velocity $u_{\tau}$:
\begin{equation}
\label{eq:v_rms}
\mathrm{u}^{+\prime}_{\mathrm{rms}} = \frac{\sqrt{(u(\mathbf{x})-\left\langle u \right\rangle)^2}}{u_{\tau}}.
\end{equation}
Spatial averaging for RMS is performed only during post-processing due to the high resolution of the simulation. In~\cref{figure: sketch} (c), the red markers represent the LBM-LES simulation results, while the blue markers correspond to the PIV experimental reference, and the black lines denote the DNS reference. Both experimental and LBM-LES results successfully capture the peak value of $u^{+\prime}_{\mathrm{rms}}$ compared with the DNS data. Beyond the peak, both LBM-LES and experimental results gradually deviate from the DNS data, which may be attributed to statistical issues and the confinement of the channel flow in the spanwise direction. Regarding the vertical velocity component $u^{+\prime}_{\mathrm{rms}}$, LBM-LES results align well with the experimental data but are slightly lower than the DNS results for $y^+ > 20$.

Considering both~\cref{figure: sketch} (b) and~\cref{figure: sketch} (c), it is evident that the LBM-LES implementation aligns closely with experimental and DNS results, demonstrating its capability to capture turbulent statistical quantities accurately. The comparison reveals that the LBM-LES method effectively replicates the key features of turbulent flows, confirming its reliability in modeling complex flow dynamics. This alignment shows the potential of LBM-LES in contributing valuable insights into the understanding of turbulence, particularly in the context of fluid dynamics simulations.
\\

\subsection{Simulation of high Reynolds number flow in a stenosis of the aorta}
\begin{figure}[H]
\includegraphics[width=\textwidth]{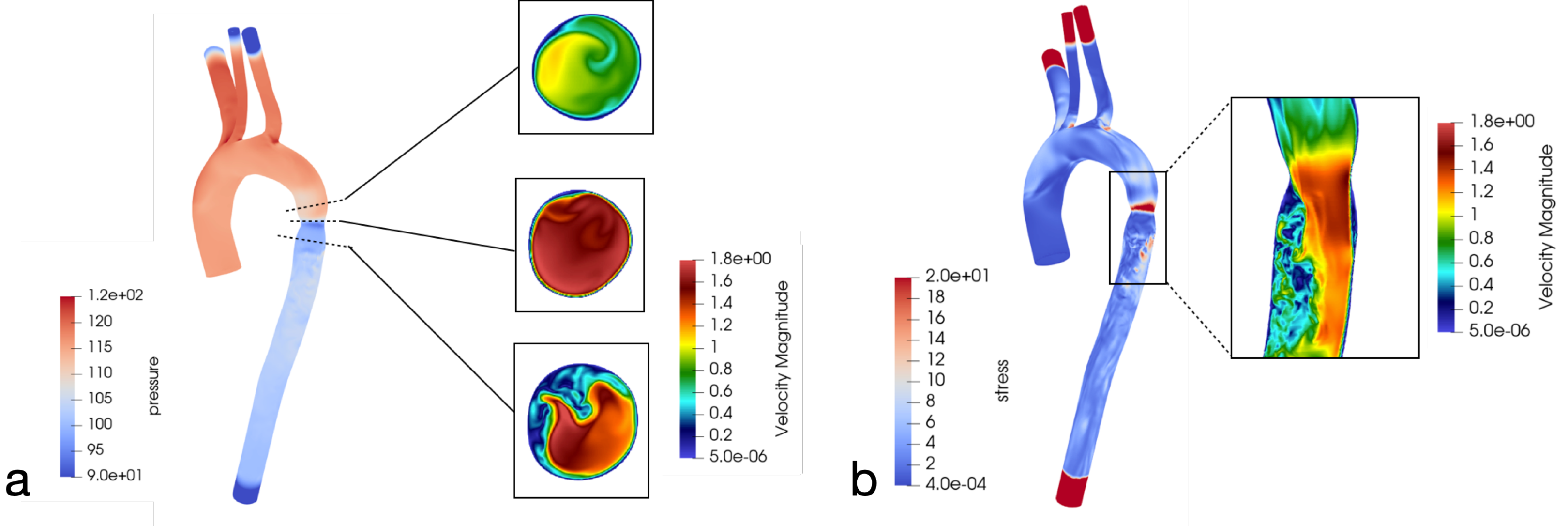}
\caption{Panel a: instantaneous snapshot of aortic stenosis simulation: on the left-hand side, the pressure distribution within the aortic flow is depicted. The right-hand side illustrates three cross-sectional views?upper, middle, and lower?proximal to the aortic constriction. This figure demonstrates the capability of HemeLB to capture the blood flow behavior of aortic flow at a stenotic location with a maximum velocity of $1.8m/s$. Panel b: The snapshot presents an instantaneous simulation of wall shear stress in stenotic aortic flow, featuring a cross-sectional visualization near the region of aortic stenosis. The velocity magnitude profile within this cross-section highlights the transition from laminar to turbulent flow, a phenomenon occurring within the stenotic segment of the aorta.} \label{figure:aorta}
\end{figure}
In this subsection, we present our capability to simulate blood flow at $Re = 4500$ utilising our LBM-LES methodology. Specifically, we apply this technique to simulate stenotic blood flow within a thoracic aorta model~\cite{Wilson2013TheResults}, which has an inlet diameter of approximately $0.01m$. The largest lattice spacing required to reasonably represent such a domain is approximately $100 \mu m$. However, the BGK collision operator faces challenges in achieving stable simulations at this resolution (especially at elevated Re) due to the limited choice of the relaxation time $\tau$. For a reliable simulation of stenotic aorta flow, a finer resolution of $20 \mu m$ for the geometry may be necessary. To illustrate the benefit of LES techniques, we set the model resolution in our study to $100 \mu m$ \R{with the Smagorinsky constant set to $C_{\mathrm{smag}} = 0.01$}. As a proof of concept, the model is initialized with an inlet velocity around $1m/s$. Pressure-free outlet conditions are established for three branches leading to the cerebral and upper limb vasculature and a singular outlet in the bottom leading to the descending aorta. The sponge regions are set near the outlets as described in~\cref{eq:sponge_visco}. As shown in~\cref{figure:aorta}, the LBM-LES based simulation is able to capture the pressure drop due to the presence of the narrowing part of the aorta. The cross-sectional views on the upper, middle, and lower section near the aortic constriction show that on the upper plane, the mean velocity is around $1m/s$, the reduced cross-section of the middle section leads to the acceleration of the blood flow to around $1.8m/s$. Furthermore, the high blood flow velocity leads to turbulent flow in the lower part of the stenotic region. The presence of the blockage will also lead to a pressure drop~\cite{feiger2020accelerating} which is observed in our simulation. We emphasise that the typical resolution for the stenotic aorta would be around $20\mu m$ resolution, whereas our configuration is able to observe similar velocity and pressure profile at $100 \mu m$ resolution.

\cref{figure:aorta} (b) provides an instantaneous snapshot illustrating the surface stress profile of the aorta, along with the velocity profile across a section near the stenotic region. This figure highlights the area proximal to the aortic constriction which exhibits significantly elevated wall shear stress, indicative of the stenosis acting as an obstruction, impeding blood flow and directing it towards the descending aorta. Consequently, the upper region of the aorta is subjected to increased blood pressure, as depicted in \cref{figure:aorta}. Elevated wall shear stress is also observed near the aortic outlets, attributed to the implementation of the sponge layer. On the right-hand side of \cref{figure:aorta} (b), the velocity profile within the aorta is presented, revealing a laminar-turbulent transition induced by aortic stenosis. Mirroring the stenotic channel flow simulation illustrated in \cref{figure: sketch} (a), the flow evolves into a turbulent state, achieving a peak velocity of approximately $1.8m/s$. This transition underscores the complexity of flow patterns in the areas of aortic constriction, reflecting the significant impact of stenosis on the hemodynamic behavior of the aorta.
\section{Uncertainty quantification visualization for the stenosis of the aorta}
In this section, we introduce a generalizable UQ visualization framework designed for LBM-based hemodynamic simulations. The goal of the framework is to provide interpretable and spatially localized representations of uncertainty in flow quantities such as pressure and stress fields over the aorta, enabling deeper insight into the sensitivity of simulation outcomes to modeling assumptions and input variability.

To demonstrate the utility of the framework, we apply it to the case of high Reynolds number flow through a stenosed aorta. The UQ framework is structured around three main components. First, we outline the end-to-end workflow that integrates ensemble simulations with statistical modeling and visualization. Second, we analyze spatial uncertainty due to variation in inflow velocity profiles. Finally, we assess uncertainty associated with different values of the Smagorinsky constant, a key parameter in LES turbulence modeling.

\subsection{Uncertain quantification visualization work flow}\label{sec:uqv}
This subsection describes the overall pipeline of the UQ visualization framework, including parameter sampling, ensemble simulation, statistical modeling of output fields, and rendering of uncertainty metrics.

\begin{figure}[H]
    \centering
    \includegraphics[width=\textwidth]{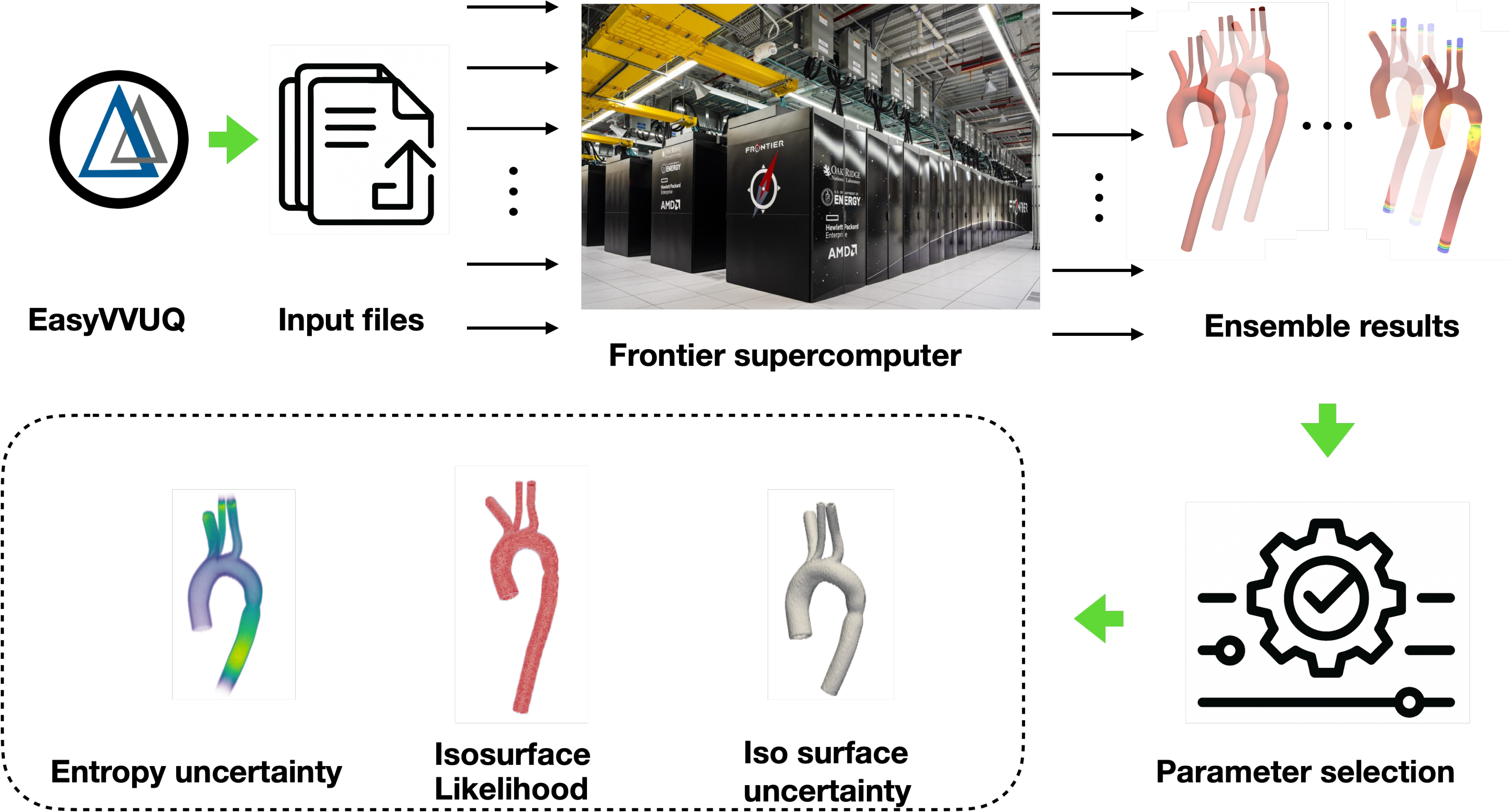}
    \caption{
        Uncertainty visualization workflow with help of EasyVVUQ and the Frontier supercomputer. Input parameter files generated by EasyVVUQ are executed as ensembles on Frontier. The resulting simulation outputs are analyzed to extract entropy, iso-surface likelihood and iso-surface uncertainties, which are then used to quantify and visualize parameter-driven uncertainty.
    }
    \label{fig:uq_workflow}
\end{figure}

Figure~\ref{fig:uq_workflow} presents a workflow for uncertainty quantification visualization with help of EasyVVUQ and UQ visualization techniques described in~\cref{subsec:uqv} in conjunction with the Frontier supercomputer. The process begins with the generation of input parameter files through EasyVVUQ, which are then used to launch an ensemble of simulations on Frontier. Each simulation corresponds to a different realization of the uncertain parameters for instances, input velocity variation and Smagorinsky constant. EasyVVUQ generates input parameter files, which are then executed as an ensemble of simulations on the Frontier supercomputer using the workflow management system FabSim3~\cite{groen2023fabsim3}. The ensemble results are post-processed to evaluate various forms of uncertainty. Three key metrics are highlighted: \textit{entropy uncertainty}, which quantifies the information-theoretic spread in simulation outputs, \textit{iso-surface likelyhood}, which defines the value of the probability where the iso-surface appears, and \textit{iso-surface uncertainty}, which captures spatial variations across isosurfaces of interest (e.g., pressure or wall stress fields). These uncertainty measures provide insight into the sensitivity of the model to input parameters and support robust decision-making in engineering and clinical applications. This workflow exemplifies how scalable UQ can be performed on exascale platforms to enable reliable and interpretable simulation outcomes. Next, we apply our framework to spatial uncertainties on the surface of the stenosis of the aorta.

\subsection{Spatial uncertainty quantification visualization on variation of inlet velocity}
To quantify the sensitivity of the flow field to uncertainties in the inlet peak velocity, we use EasyVVUQ~\cite{richardson2020easyvvuq} to construct a non-intrusive Polynomial Chaos Expansion (PCE) surrogate. The uncertain parameter is the maximum inlet velocity from heart $v_{\max}$ (in m\,s$^{-1}$), applied as a scalar multiplier to a normalized inlet velocity waveform $\hat{u}(\mathbf{x}, t)$ obtained from physiological measurements. The inlet profile for each realization is given by:
\begin{equation}
    u(\mathbf{x}, t; v_{\max}) = v_{\max} \, \hat{u}(\mathbf{x}, t), 
    \quad v_{\max} \in [0.4, \, 1.1] .
\end{equation}
We assume a uniform prior distribution for $v_{\max}$ to reflect typical clinical variability. The PCESampler is configured with a total polynomial order $p_o=2$ to generate a minimal set of quadrature nodes (Gauss?Legendre rule) consistent with the input distribution. Each quadrature node corresponds to a distinct HemeLB simulation, automatically prepared using an EasyVVUQ encoder that modifies the inlet boundary condition file according to $v_{\max}$.

\begin{figure}[H]
\begin{adjustbox}{center}
\includegraphics[width=1.2\textwidth]{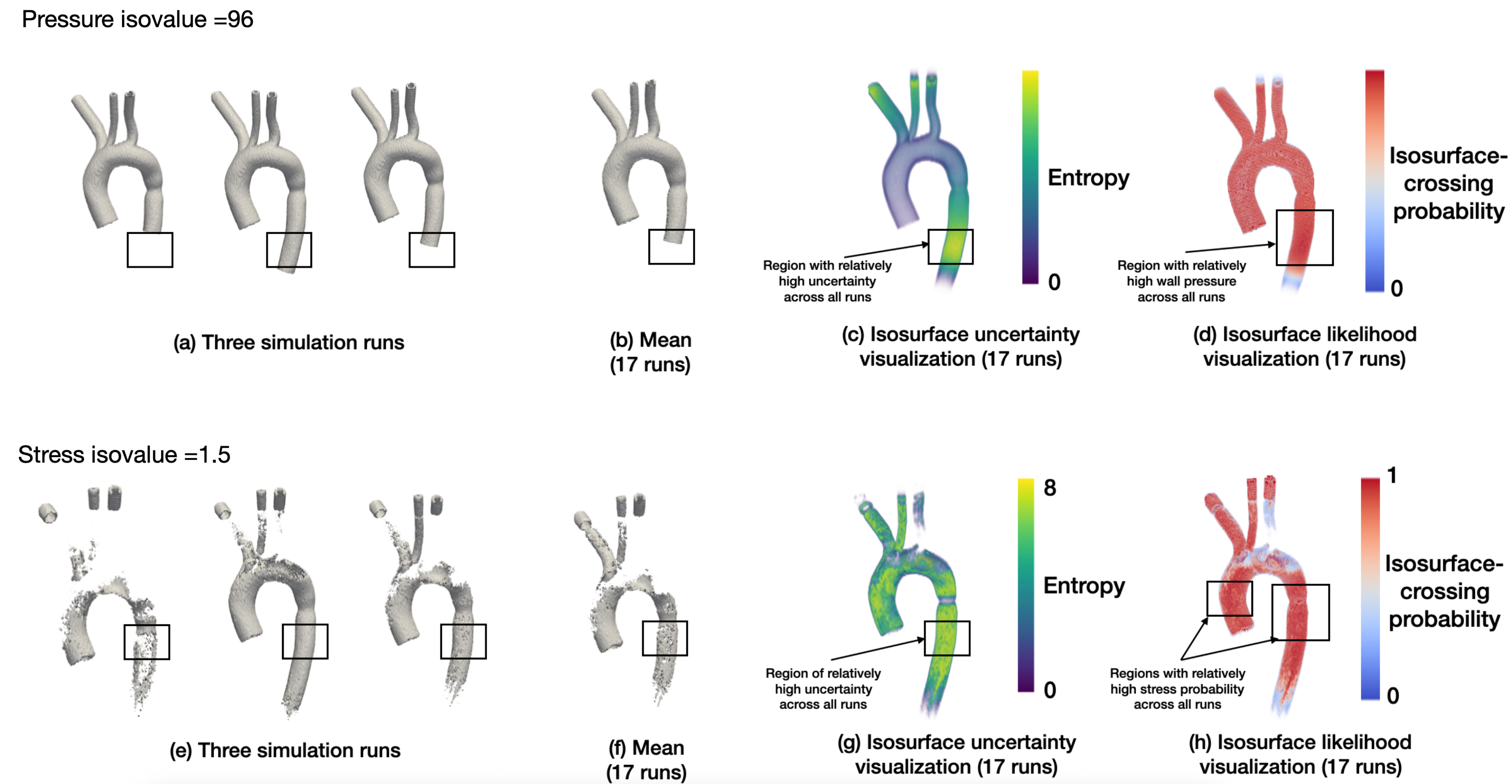}
\end{adjustbox}
\caption{Uncertainty visualization for (top row) pressure isovalue $p=96$ mmHg and (bottom row) wall shear stress isovalue $\tau_{w}=1.5$. 
    (a,e) Isosurfaces from three representative runs out of a 17-run ensemble, showing noticeable positional variation across runs. 
    (b,f) Mean isosurface computed over all runs, capturing the average geometry but not local variability. 
    (c,g) Entropy of the isosurface topology distribution, with yellow indicating regions of high positional uncertainty and blue indicating stable regions across runs. 
    For the pressure field, entropy is low upstream of the stenosis and high downstream, reflecting the transition from laminar to turbulent flow. 
    For WSS, elevated entropy regions occur primarily in high-shear zones downstream of the stenosis. 
    (d,h) Isosurface-crossing probability, with red indicating the most probable locations of the isosurface and blue the least probable, highlighting spatial consistency in isosurface positions across the ensemble.}
    \label{figure:uncertainty_vis_stress_isosurface}
\end{figure}

\begin{figure}[H]
\centering
\begin{adjustbox}{center}
\includegraphics[width=1.1\textwidth]{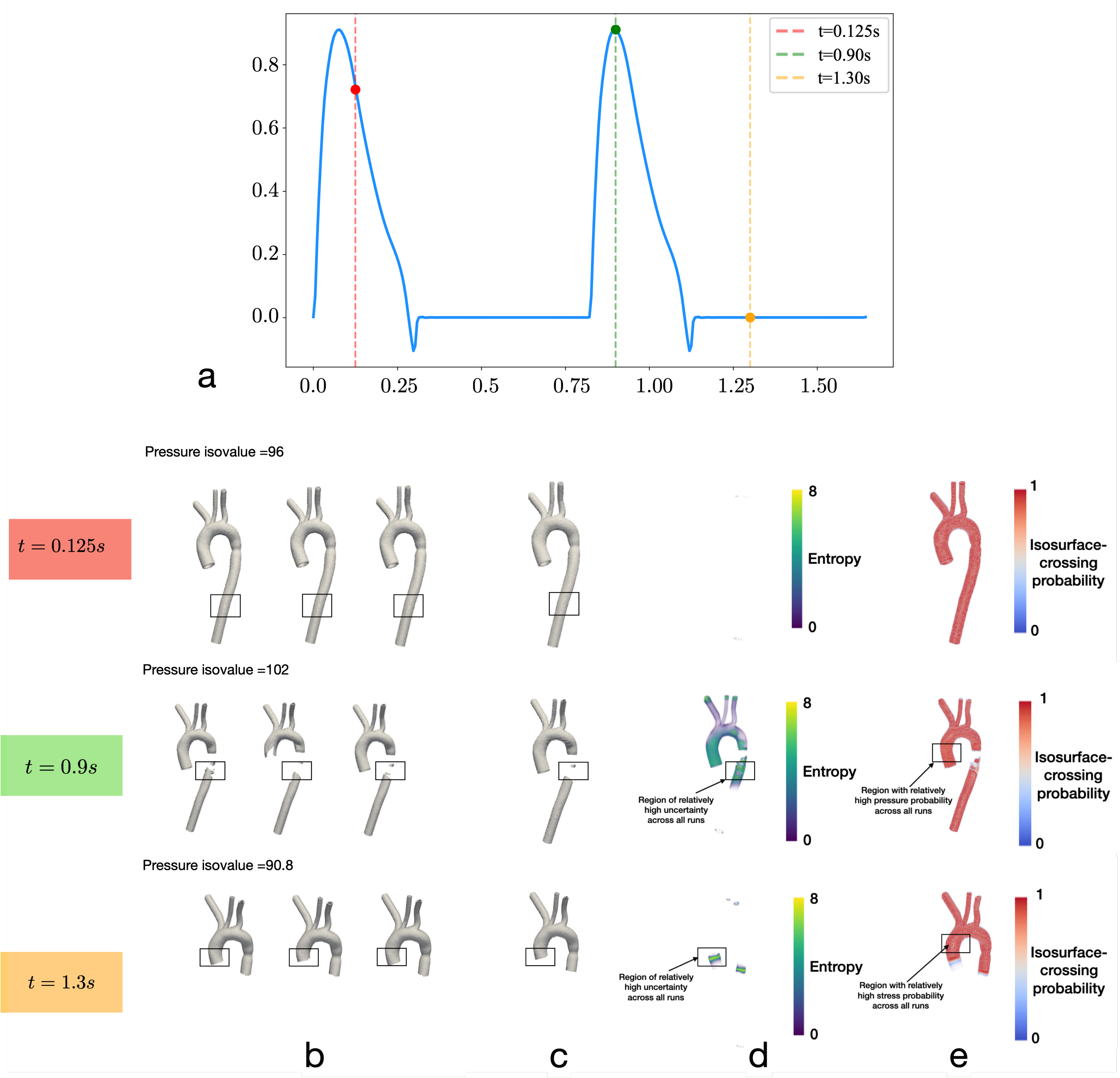}
\end{adjustbox}
\caption{Uncertainty visualization of the pressure field for an ensemble of 6 runs with varying Smagorinsky constants.
Heartbeat profile showing three selected time instants: $t=0.125\mathrm{s}$ (after the first peak, during the descending phase), $t=0.9\mathrm{s}$ (at the peak), and $t=1.3\mathrm{s}$ (late diastole).
(b) Isosurfaces of pressure for the chosen isovalue at three example runs. (c) Mean isosurface computed from all six runs in the ensemble. (d) Entropy field, where yellow indicates regions with high positional variability of the isosurface across the ensemble. (e) Isosurface probability, where red indicates highly probable positions of the isosurface and blue indicates low-probability positions.} \label{figure:uncertainty_smag_isosurface}
\end{figure}

The resulting ensemble consists of 17 high-resolution simulations, executed on the Frontier exascale supercomputer. \R{Each run was performed on 20 GPU-accelerated nodes, with each node equipped with 4 $\times$ AMD MI250X GPUs. The simulation time is 1 hour with resolution of 100 $\mu$m and the heart beat profile is depicted in Fig.~\ref{figure:uncertainty_smag_isosurface} (a).} Figure~\ref{figure:uncertainty_vis_stress_isosurface} (a) shows the pressure isosurface ($p=96$ mmHg)  from three representative runs, revealing noticeable positional variation of the isosurface caused by differences in $v_{\max}$. As direct visual comparison becomes increasingly impractical with larger ensembles, we apply our spatial uncertainty metrics to the full set of runs. Thus, we compute and visualize the mean, entropy, and isosurface-crossing probability metrics in Fig.~\ref{figure:uncertainty_vis_stress_isosurface}(b)-(d). These metrics provide a concise summary of isosurface positional variations and, importantly, are independent of the number of ensemble runs.

Figure~\ref{figure:uncertainty_vis_stress_isosurface}(b) shows the isosurface obtained from the mean of all $17$ runs. While this average shape captures the overall geometry, it does not convey which regions are more or less uncertain across the ensemble. To address this, Fig.~\ref{figure:uncertainty_vis_stress_isosurface}(c) visualizes the \emph{entropy} metric, where yellow indicates regions of high positional variability and blue marks more stable regions. The colormap reveals that upstream of the stenosis, the pressure field exhibits relatively low uncertainty, as the flow remains largely laminar and insensitive to inlet velocity variation. In contrast, high-entropy regions emerge downstream of the stenosis, where the accelerated jet transitions to turbulence and flow structures become highly sensitive to inlet conditions. 

Finally, Fig.~\ref{figure:uncertainty_vis_stress_isosurface}(d) presents the isosurface-crossing probability, with red highlighting the most probable isosurface positions and blue indicating less likely positions. This probability map complements the entropy visualization by showing where the isosurface location is most consistently observed across the ensemble. Figure~\ref{figure:uncertainty_vis_stress_isosurface}(e)-(h) summarizes the spatial uncertainty analysis for wall shear stress fields under inlet velocity variability. The wall shear stress can be defined as:
\begin{equation}
    \tau_w = \mu \left. \frac{\partial u_t}{\partial n} \right|_{\text{wall}},
\end{equation}
where $\mu$ is the dynamic viscosity, $u_t$ is the tangential velocity component at the wall, and $n$ is the wall-normal coordinate. Figure~\ref{figure:uncertainty_vis_stress_isosurface} (e) presents isosurfaces from three representative runs, illustrating substantial positional variation due to differences in $v_{\max}$. Panels (f) show the mean isosurface from all 17 runs, which captures the average shape but conceals local variability. Panels (g) display the entropy metric, where yellow regions denote high positional uncertainty and blue regions indicate stability. For the pressure isosurface ($p=96$ mmHg), entropy is low upstream of the stenosis?where the flow remains laminar and insensitive to inlet velocity changes?but increases sharply downstream, where the jet accelerates, transitions to turbulence, and becomes highly sensitive to initial conditions. In contrast, for the wall shear stress isosurface ($\tau_w = 1.5$), elevated entropy is concentrated in regions of intense shear downstream of the stenosis, where turbulent fluctuations induce significant spatial uncertainties. Within the stenotic throat itself, entropy remains low, indicating high certainty in the presence and location of a high-stress region - an unambiguous marker of the pathological constriction. Finally, panel (h) shows the isosurface-crossing probability, with red indicating the most probable locations and blue indicating the least probable. This metric complements the entropy maps by pinpointing regions of consistent isosurface occurrence across the ensemble, aiding interpretation of the clinical relevance of WSS variability. 

Our uncertainty visualization framework extends beyond traditional one-dimensional quantitative plots by providing spatially resolved, three-dimensional depictions of variability in clinically relevant flow metrics such as pressure and wall shear stress. By mapping statistical measures,  including entropy and isosurface-crossing probability, directly onto patient-specific vascular geometries, the framework offers an intuitive representation of where predictions are most certain or uncertain. This visual context allows clinicians to quickly identify regions of hemodynamic instability (e.g., turbulent zones downstream of a stenosis) and confidently interpret stable high-risk areas (e.g., consistently elevated WSS at the stenotic throat). Such visualizations reduce the cognitive barrier for non-expert users, who may not be familiar with interpreting statistical curves or sensitivity indices, and support decision-making by linking uncertainty directly to anatomical landmarks. In doing so, the method bridges the gap between advanced computational uncertainty quantification methods and practical, immediately accessible, and explainable clinical insight.
\subsection{Spatial uncertainty quantification visualization on variation of Smagorinsky constant}
For the Smagorinsky constant study, we employed EasyVVUQ to build a non-intrusive Polynomial Chaos Expansion surrogate. The uncertain parameter is the Smagorinsky constant $C_{\text{smag}}$, which controls the modeled subgrid-scale viscosity in LES. We assigned $C_{\text{smag}}$ a uniform distribution in the range [0.01, 1.0], spanning values from typical literature defaults to significantly larger turbulence model values.

The PCE Sampler was configured with a total polynomial order of 2, producing the minimal set of quadrature nodes required for the PCE approximation while ensuring compatibility with the uniform input distribution through a Gauss?Legendre quadrature rule. Each quadrature node was mapped to a HemeLB simulation case using a custom EasyVVUQ encoder that updates the turbulence model configuration file with the sampled $C_{\text{smag}}$ value. The resulting set of 6 simulations was executed on the Frontier supercomputer, and the outputs were post-processed to extract pressure and wall shear stress fields for subsequent spatial uncertainty visualization.
Figure~\ref{figure:uncertainty_smag_isosurface} summarizes the uncertainty visualization for the pressure field across an ensemble of 6 runs in which the Smagorinsky constant was varied. Figure~\ref{figure:uncertainty_smag_isosurface} (a) shows the heartbeat profile from which three representative time instants were selected: $t=0.125\mathrm{s}$ (after the first peak, during the descending phase), $t=0.9\mathrm{s}$ (at the second peak), and $t=1.3\mathrm{s}$ (during the later stabilised period). For each time point, Fig.~\ref{figure:uncertainty_smag_isosurface} (b) presents the pressure isosurfaces from three example simulations, illustrating the variability in spatial position due to changes in the Smagorinsky constant. In Fig.~\ref{figure:uncertainty_smag_isosurface} (c) shows the mean isosurface computed over all runs, which alone does not reveal regions of greater or lesser positional stability.

The entropy visualizations in Fig.\ref{figure:uncertainty_smag_isosurface}~(d) display the entropy field, where yellow indicates regions with large positional fluctuations of the isosurface across the ensemble, while blue indicates stable locations. Immediately after the first peak ($t = 0.125\mathrm{s}$), entropy remains low throughout the domain, indicating that uncertainty in the Smagorinsky constant exert little influence during the early stages of the simulation. At the second peak ($t=0.9\mathrm{s}$), high entropy zones extend both upstream and downstream of the constriction, indicating that strong acceleration amplifies Smagorinsky-dependent variability throughout the domain. During the later stabilized period ($t=1.3\mathrm{s}$), entropy decreases markedly, and high-probability zones become more spatially coherent, suggesting that in steady or quasi-steady flow conditions, the effect of varying the Smagorinsky constant is reduced. Notably, entropy in the stenotic throat is always low, indicating high certainty in the presence and location of a consistently high-stress region?an unambiguous indicator of pathological significance. Figure\ref{figure:uncertainty_smag_isosurface}~(e) presents the isosurface-crossing probability, which complements the entropy analysis by highlighting the most probable surface positions in red and the least probable in blue. 

\begin{figure}[H]
\centering
\begin{adjustbox}{center}
  \includegraphics[width=1.2\textwidth]{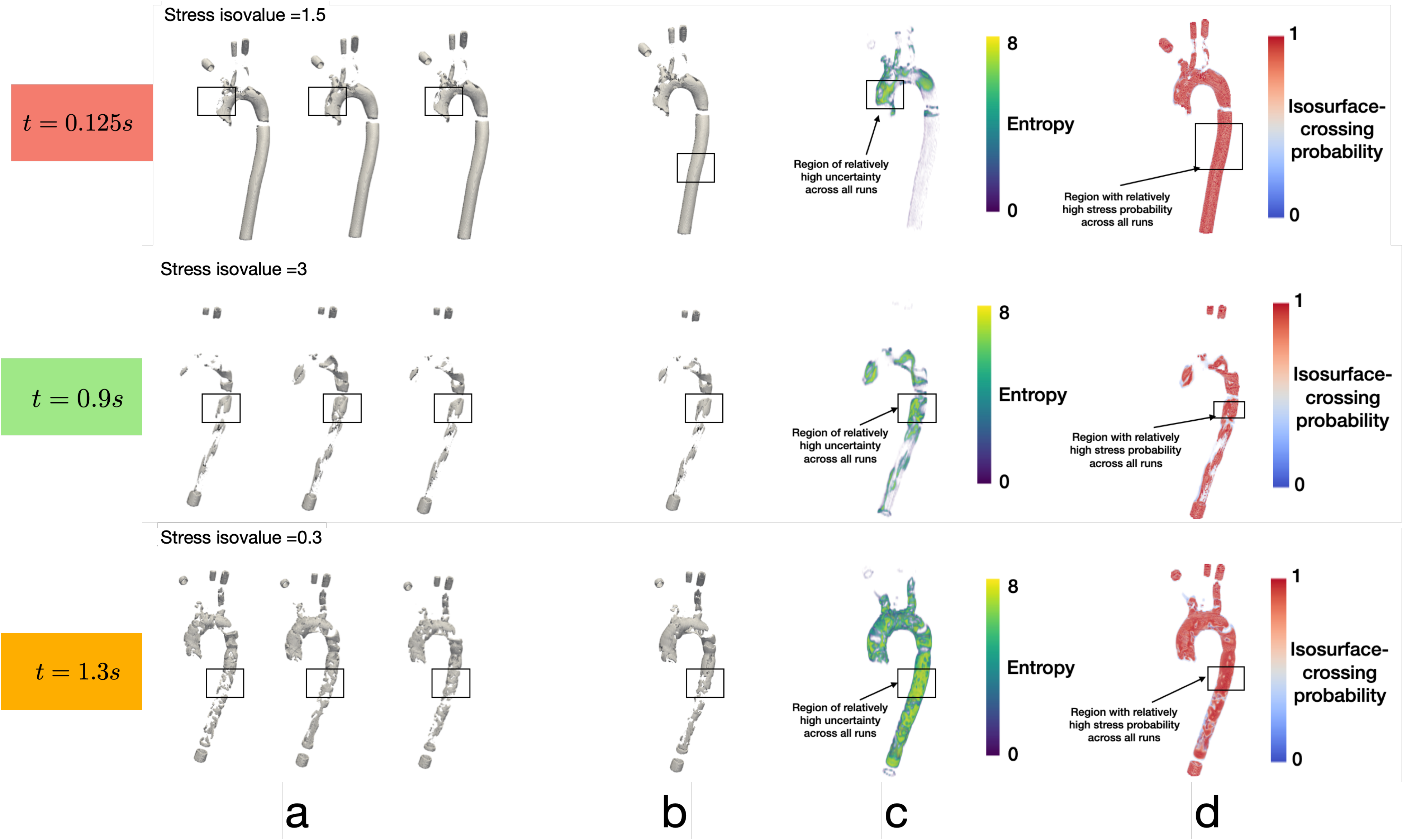}
\end{adjustbox}
\caption{Uncertainty visualization of WSS fields for an ensemble of 6 runs with varying Smagorinsky constants. (a) Isosurfaces of WSS for three example simulations at each selected time instant: $t=0.125\mathrm{s}$ (after the first peak, during the descending phase), $t=0.9\mathrm{s}$ (at the second peak), and $t=1.3\mathrm{s}$ (later stabilised period). The stress isovalue is 1.5, 3, and 0.3, respectively. (b) Mean WSS isosurface computed from all runs. (c) Entropy field, where yellow indicates regions with high positional variability of the WSS isosurface across the ensemble. (d) Isosurface-crossing probability, where red indicates highly probable WSS isosurface positions and blue indicates low-probability positions.} \label{figure:uncertainty_smag_isosurface_1.3}
\end{figure}
To complement the uncertainty analysis, Figure~\ref{figure:uncertainty_smag_isosurface_1.3} presents spatially resolved uncertainty in the WSS field across an ensemble of 6 runs with varying Smagorinsky constants. The three time instants?$t=0.125\mathrm{s}$ (after the first peak, descending phase), $t=0.9\mathrm{s}$ (at the second peak), and $t=1.3\mathrm{s}$ (later stabilised period) - capture key dynamic phases of the cardiac cycle. For each instant, the WSS isosurface values were chosen as 1.5, 3, and 0.3, respectively, to reflect representative stress thresholds.
Figure~\ref{figure:uncertainty_smag_isosurface_1.3} (a) show WSS isosurfaces from three representative runs, demonstrating noticeable positional variation depending on the Smagorinsky constant. Figure~\ref{figure:uncertainty_smag_isosurface_1.3} (b) presents the mean WSS isosurfaces across all ensemble members, which mask the underlying variability. Panels (c) depict entropy distributions, with yellow regions highlighting zones of high WSS positional variability. Notably, at $t=0.125\mathrm{s}$ and $t=0.9\mathrm{s}$, elevated entropy is concentrated downstream of the stenosis, indicating strong sensitivity of turbulent shear stresses to the SGS parameter during unsteady acceleration and deceleration phases. At $t=1.3\mathrm{s}$, the entropy field shows a more uniform and lower magnitude, reflecting fully developed chaos due to the change of the Smagorinsky constant. Figure~\ref{figure:uncertainty_smag_isosurface_1.3} (d) shows the isosurface-crossing probability, where red indicates locations consistently intersected by the WSS isosurface across the ensemble, and blue marks low-likelihood positions. These probability maps confirm that high WSS zones near the stenosis remain spatially consistent, while downstream regions exhibit substantial variability when the Smagorinsky constant changes.

Overall, the WSS analysis reveals that $C_{\text{smag}}$ parameter variation has the greatest influence during highly dynamic flow phases?particularly during early post-peak deceleration and peak systolic acceleration?while its impact diminishes during the later stabilised phase. This pattern parallels the pressure-field uncertainty trends but is more spatially localized near the vessel walls, underscoring the distinct physical sensitivity of wall shear stress to turbulence-model parameters.

\section{Conclusion}\label{sec:conclusion}
In this study, we presented a flexible uncertainty quantification and visualization framework for lattice Boltzmann method simulations of vascular hemodynamics, demonstrated on high?Reynolds number flow through a stenosed aorta. The framework combines EasyVVUQ-based parameter sampling with spatially resolved three-dimensional uncertainty metrics, enabling anatomically contextualized assessments of clinically relevant quantities such as pressure and wall shear stress (WSS). Two case studies were examined: variation in inlet peak velocity and variation in the Smagorinsky constant $C_{\mathrm{smag}}$ used in large-eddy simulation turbulence modeling. Variation in inlet velocity produced pronounced spatial uncertainty downstream of the stenosis, particularly in regions undergoing laminar?turbulent transition, while maintaining low uncertainty upstream. In contrast, $C_{\mathrm{smag}}$ variation induced relatively small changes in the large-scale pressure field but generated localized increases in WSS uncertainty within high-shear regions downstream of the stenosis. The temporal analysis revealed that the influence of $C_{\mathrm{smag}}$ is phase-dependent: uncertainty is highest during dynamic phases of the cardiac cycle (post-peak deceleration and peak systolic acceleration) and lowest during later stabilised flow. In both parameter-variation cases, the consistently low entropy at the stenotic throat indicated a robust high-stress region, suggesting a reliable marker for pathological assessment. By mapping entropy and isosurface-crossing probability directly onto patient-specific geometries, the proposed framework surpasses conventional one-dimensional statistical plots, offering immediately interpretable, spatially localized insights into model sensitivity. This capability is particularly valuable for non-expert users, as it allows direct linkage between simulation variability and anatomical structures of clinical interest. Although demonstrated here for aortic stenosis, the methodology is broadly applicable to other vascular domains and flow regimes, supporting the integration of uncertainty-aware modeling into patient-specific simulations and advancing the use of computational hemodynamics in future clinical workflows.

\section{Acknowledgements}
We acknowledge funding support from European Commission CompBioMed Centre of Excellence (Grant No. 675451 and 823712). Support from the UK Engineering and Physical Sciences Research Council under the projects ``UK Consortium on Mesoscale Engineering Sciences (UKCOMES)" (Grant No.
EP/R029598/1) and ``Software Environment for Actionable and VVUQ-evaluated Exascale Applications (SEAVEA)" (Grant No. EP/W007711/1) is gratefully acknowledged. 
This research used resources of the Oak Ridge Leadership Computing Facility at Oak Ridge National Laboratory, which is supported by the Office of Science of the U.S. Department of Energy under Contract No. DE-AC05-00OR22725. It also benefited from the 2024?2025 DOE INCITE award under the project ``COMPBIO3", which provided access to supercomputing resources at both the Oak Ridge and Argonne Leadership Computing Facilities.
%
%
%
\bibliographystyle{splncs04}
\bibliography{ref_Mendeley}
\end{document}